\documentclass{pasj01}
\onecolumn 
\Received{$\langle$reception date$\rangle$}
\Accepted{$\langle$acception date$\rangle$}
\Published{$\langle$publication date$\rangle$}
\begin{document}

\title{KOOLS--IFU: Kyoto Okayama Optical Low-dispersion Spectrograph with Optical-Fiber Integral Field Unit}
\author{Kazuya \textsc{Matsubayashi}\altaffilmark{1},
 Kouji \textsc{Ohta}\altaffilmark{2},
 Fumihide \textsc{Iwamuro}\altaffilmark{2},
 Ikuru \textsc{Iwata}\altaffilmark{3},
 Eiji \textsc{Kambe}\altaffilmark{3},
 Hironori \textsc{Tsutsui}\altaffilmark{4},
 Hideyuki \textsc{Izumiura}\altaffilmark{4}, 
 Michitoshi \textsc{Yoshida}\altaffilmark{3}, and
 Takashi \textsc{Hattori}\altaffilmark{3}
}%
\altaffiltext{1}{Okayama Observatory, Kyoto University, 3037-5 Honjo, Kamogata-cho, Asakuchi, Okayama 719-0232, Japan}
\altaffiltext{2}{Department of Astronomy, Kyoto University}
\altaffiltext{3}{Subaru Telescope, National Astronomical Observatory of Japan}
\altaffiltext{4}{Okayama Branch Office, National Astronomical Observatory of Japan}
\email{kazuya@kusastro.kyoto-u.ac.jp}


\maketitle

\begin{abstract}

Observations of transient objects, such as short gamma-ray bursts (GRBs) and electromagnetic counterparts of gravitational wave sources, require prompt spectroscopy.
To carry out prompt spectroscopy, we developed an optical-fiber integral field unit (IFU) and connected it with an existing optical spectrograph KOOLS.
KOOLS with IFU, hereafter KOOLS--IFU, was mounted on the Okayama Astrophysical Observatory 188-cm telescope.
The fiber core and cladding diameters of the fiber bundle are 100 \micron\, and 125 \micron, respectively, and 127 fibers are hexagonally close-packed in the sleeve of the two-dimensional (2D) fiber array.
We conducted test observations to measure the KOOLS--IFU performance and obtained the following conclusions:
(1) the spatial sampling is \timeform{2''.34} $\pm$ \timeform{0''.05} per fiber, and the total field of view (FoV) is \timeform{30''.4} $\pm$ \timeform{0''.65} with 127 fibers;
(2) the observable wavelength and the spectral resolving power of the grisms of KOOLS are 4030--7310 \AA\, and 400--600; 5020--8830 \AA\, and 600--900; 4160--6000 \AA\, and 1000--1200; and 6150--7930 \AA\, and 1800--2400, respectively; and
(3) the estimated limiting magnitude is 18.2--18.7 AB mag during 30-min exposure under the optimal condition.

\end{abstract}

\section{Introduction}

Optical wavelength high-cadence surveys of transient objects have been on the rise.
The intermediate Palomar Transient Factory (iPTF) employed a wide-field CCD camera attached to the 48-inch Samuel Oschin Schmidt Telescope at Palomar Observatory and conducted a wide-field survey with a 90-s to 5-day cadence \citep{Rau:2009,Law:2009}.
All-Sky Automated Survey for Supernovae (ASAS-SN) observes the whole sky every night with 20 optical telescopes to find bright supernovae \citep{Shappee:2014}.
The transient surveys, such as the Zwicky Transient Facility (ZTF) and the Tomo-e Gozen surveys with Kiso 1.0-m Schmidt telescope \citep{Sako:2016}, plan to cover most of the sky visible from these facilities using about a one-hour cadence.
X-ray and gamma-ray high cadence all sky surveys are also ongoing (e.g., Swift, Fermi, and MAXI).
Among them, Swift \citep{Gehrels:2004} is capable of detecting a gamma-ray burst (GRB) typically in only a minute, and if its X-Ray Telescope (XRT) detects the GRB's X-ray afterglow, it sends off an alert of detection immediately.
In addition to electromagnetic waves, transient objects can also be detected via new messengers of gravitational wave (GW; \cite{Abbott:2016}) and high-energy neutrino \citep{IceCube:2018}.
Facilities of the new messengers also promptly send alerts of detection.

The surveys and facilities using these new messengers have been detecting various interesting transients.
To unveil the nature of the detected transients, prompt follow-up observations are important, particularly optical spectroscopic observations, within a few minutes after detections.
However, such prompt observations are very challenging.
Spectroscopic observations generally require more photons to attain adequate signal-to-noise ratios when compared with that required in the imaging observations and can be optimally performed using large telescopes.
Slit spectroscopy is a simple and general method for conducting spectroscopic observations.
However, it is difficult to make observations promptly as aligning the target with the slit properly is time consuming, particularly if the target is faint.
This is because generally the telescope pointing accuracy is not accurate enough, and a longer exposure time is required for target acquisition.

An optical-fiber integral field spectrograph (e.g., \cite{Ren:2002}) is a good tool to use for prompt follow-up spectroscopy.
Various integral field units have been developed.
One utilizing a fiber bundle can be compact compared with other types because of the fiber flexibility and the simple optical layout.
Just inserting the unit into the light path of the telescope in front of the entrance aperture of the existing instrument is all that is needed to use integral field spectroscopy (IFS) to make a spectroscopic measurement of the sky over a two-dimensional field of view (FoV).
Development of an insertable and retractable system of such a unit is not difficult and has only a small impact on the existing instruments.
IFS is a powerful tool for prompt follow-up spectroscopy because its exposure can be started just after the completion of the telescope pointing without slit acquisitions if the FoV of the IFS is larger than the combination of the telescope pointing error and the positional error of the target object.

To carry out prompt follow-up spectroscopy for transients, we developed a prototype of an optical-fiber integral field unit (IFU).
In our prototype, one end of the fiber bundle is located at the focal plane of the telescope and the other end is fed to an existing spectrograph---the Kyoto Okayama Optical Low-dispersion Spectrograph KOOLS \citep{Yoshida:2005}.
In the focal plane, individual fibers are placed in a closely packed arrangement, while at the spectrograph entrance they are linearly aligned.
KOOLS with IFU, hereafter referred to as KOOLS--IFU, was attached to the Okayama Astrophysical Observatory (OAO) 188-cm telescope and observed some transients and spatially extended objects.
This paper describes the performance of KOOLS--IFU and the results of the test observations with the OAO 188 cm telescope.
The design and the components of KOOLS--IFU are described in section \ref{sec:instrument}.
Section \ref{sec:test-obs} describes the test observation condition and data reduction.
Section \ref{sec:result} presents the results of the KOOLS--IFU performance derived from the data obtained through daytime tests and test observations.
The summary and future plans are given in section \ref{sec:summary}.

\section{Instrument}
\label{sec:instrument}

\subsection{Existing Instruments}
\label{sec:existing-inst}

\begin{figure}
\begin{center}
\includegraphics[width=160mm]{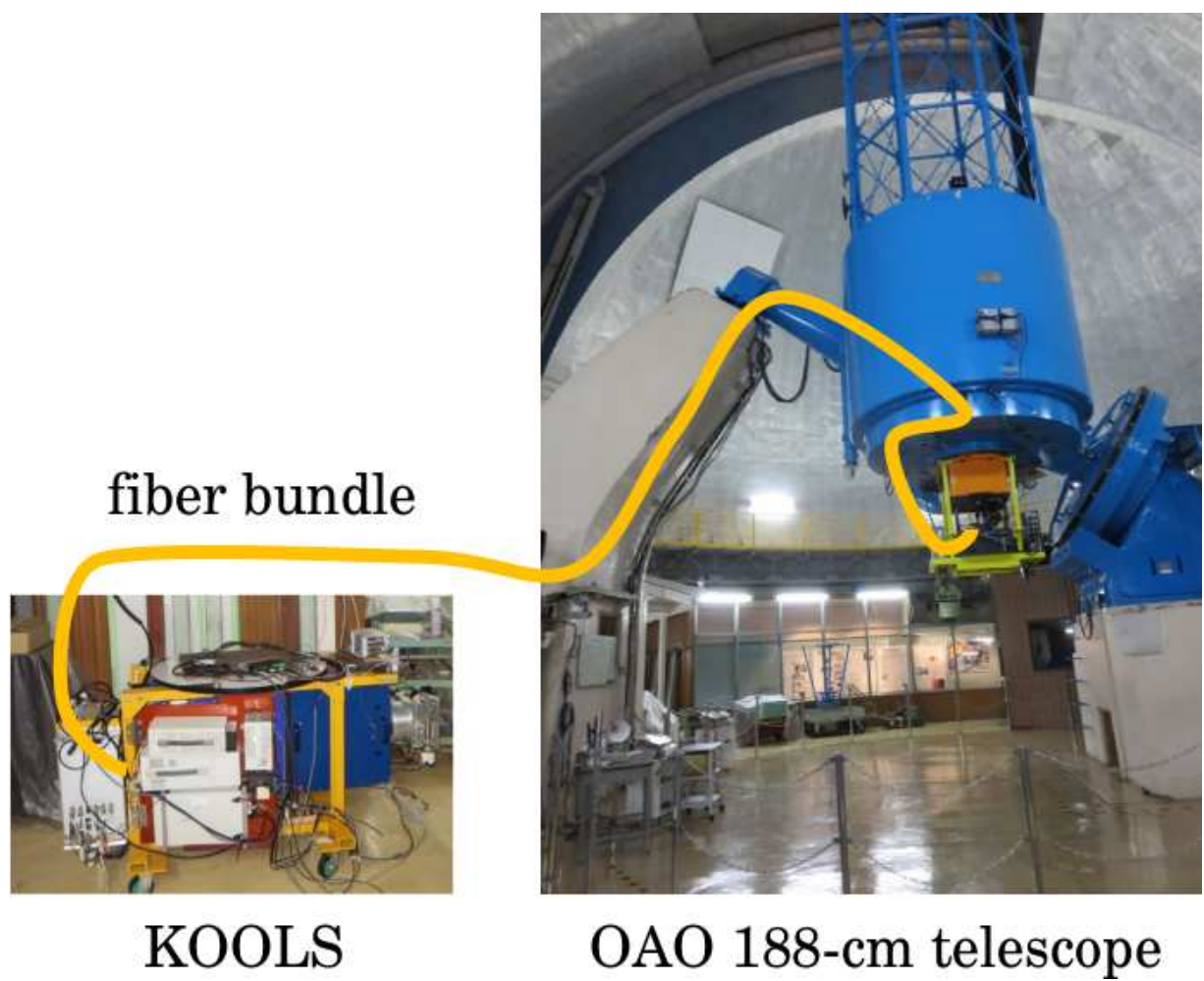}
\end{center}
\caption{Overall view of KOOLS--IFU.
The left and right panels show pictures of KOOLS and the OAO 188-cm telescope with the Cassegrain unit of HIDES-F, respectively.
North is to the left on the right panel.
The input and output parts of the IFU are mounted on the HIDES-F Cassegrain unit and KOOLS, respectively.
The fiber bundle is shown with a yellow line and is fixed on the telescope structures.
(Color online)
}
\label{fig:kools-ifu-overall}
\end{figure}

Figure \ref{fig:kools-ifu-overall} shows an overall view of KOOLS--IFU.
KOOLS--IFU makes use of the OAO 188-cm telescope, HIDES-F, and KOOLS.
The F-ratio of the Cassegrain focus of the telescope is 18.
HIDES-F is a fiber-fed optical high dispersion spectrograph for this telescope \citep{Izumiura:1999,Kambe:2013}.
The light is collected at the Cassegrain focus and is fed by fibers to the echelle spectrograph located at the coud\'e focus.
The IFU input part of KOOLS--IFU is installed at the Cassegrain unit of HIDES-F.
The fiber bundle is fed to the KOOLS spectrograph on the observing floor near the north pier of the telescope.

\begin{figure}
\begin{center}
\includegraphics[width=160mm]{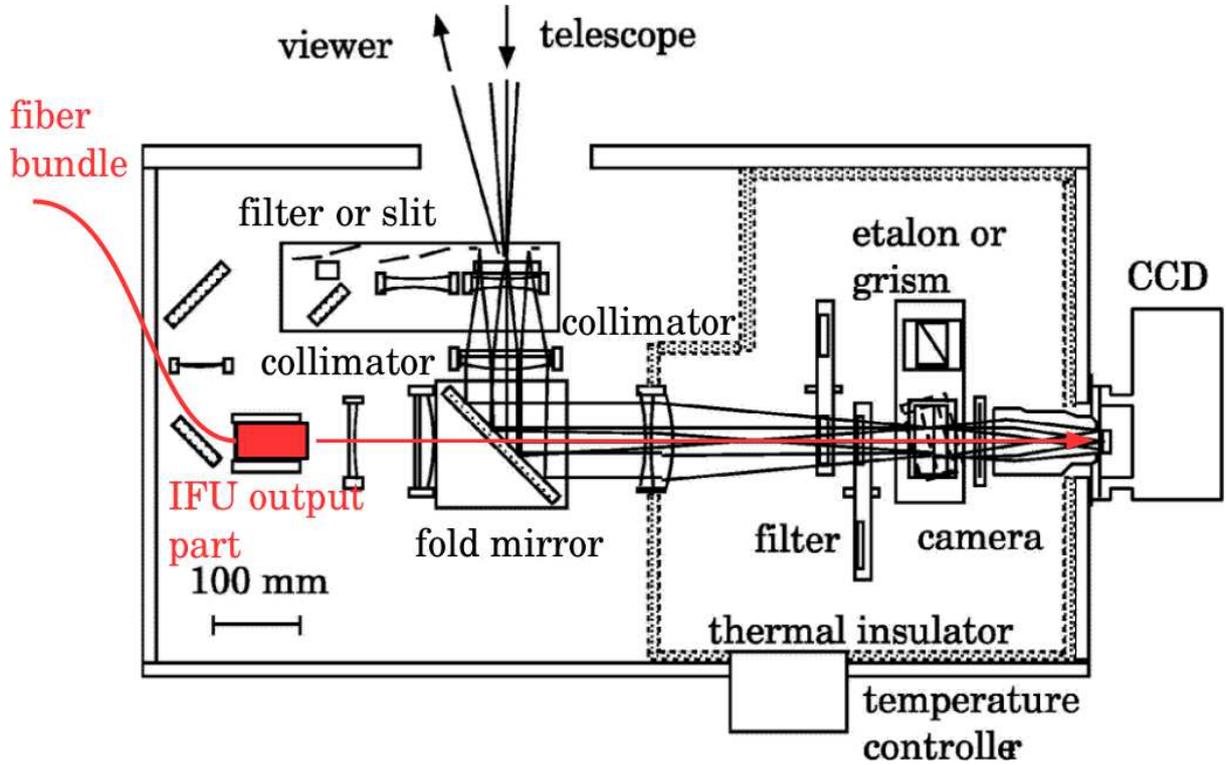}
\end{center}
\caption{KOOLS and KOOLS--IFU optical layout.
When KOOLS is mounted on the Cassegrain focus of the OAO 188-cm telescope, light from the telescope enters from top of this image.
In the case of KOOLS--IFU observations, the fold mirror is retracted and telescope light comes from the IFU output part.
The thermal insulator and the temperature controller are formerly utilized.
(Color online)
}
\label{fig:kools-layout-fiber}
\end{figure}

KOOLS is a renewal spectrograph of the multimode optical instrument Kyoto 3DI \citep{Ohtani:1998, Ishigaki:2004} mounted on the Cassegrain focus of the OAO 188-cm telescope.
Figure \ref{fig:kools-layout-fiber} shows the KOOLS optical layout and light paths \citep{Yoshida:2005}.
When KOOLS is mounted on the Cassegrain focus of the OAO 188-cm telescope, light from the telescope enters from top of this figure, and the light path to the CCD is shown with black lines.
In the case of KOOLS--IFU observations, the IFU output part is mounted left of the collimator, and the fold mirror is retracted downward so as not to interrupt light from the fibers.
The light path of this mode is shown as a red line.
Tables \ref{tb:param-kools} and \ref{tb:param-grism} show the parameters of KOOLS and its grisms\footnote{$<$http://www.oao.nao.ac.jp/\~{}kools/spec/index.html$>$}.
Three grisms can be mounted on the grism stage.
An order-sorting filter, Y49, which blocks light with wavelengths shorter than $\sim$4900 \AA, is usually used for observations with the No. 2 and VPH 683 grisms to block the second-order light.

\begin{table}
\tbl{KOOLS parameters}{
\begin{tabular}{cc}
\hline
CCD & SITe 2K $\times$ 4K$^a$ \\
CCD pixel size & 15 \micron \\
CCD gain & 2.25 electron ADU$^{-1}$ \\
CCD readout noise & 10--22.5 electrons pixel$^{-1}$ \\
Focal length of collimator & 300 mm \\
Acceptable focal plane size of collimator & 60 mm \\
Acceptable F-ratio of collimator & 8.6 \\
Focal length of camera lens & 81 mm \\
\hline
\end{tabular}}
\label{tb:param-kools}
\begin{tabnote}
a: Controlled by Messia-V system \citep{Nakaya:2004} \\
\end{tabnote}
\end{table}

\begin{table}
\tbl{Parameters of KOOLS grisms}{
\begin{tabular}{ccccc}
\hline
Grism & No. 5 & No. 2 & VPH 495$^a$ & VPH 683$^a$ \\
\hline
Wavelength (\AA) & 4000--7400 & 5700--8500 & 4500--5400 & 6200--7200 \\
Spectral resolution$^b$ & 400--600 & 600--800 & $\sim$1100 & $\sim$1100 \\
Spectral sampling (\AA\, pixel$^{-1}$)$^c$ & 4.88 & 3.78 & 1.78 & 2.24 \\
\hline
\end{tabular}}
\label{tb:param-grism}
\begin{tabnote}
a: Volume phase holographic grism \citep{Ebizuka:2011a,Ebizuka:2011b} \\
b: With \timeform{1''.8} (= 297 \micron) slit \\
c: After 2-pixel binning \\
\end{tabnote}
\end{table}

\subsection{Optical Design}
\label{sec:optical-design}

One of our main scientific goals is prompt optical follow-up spectroscopy of short GRBs. 
Their typical positional error by Swift/XRT is $\sim$\timeform{5''} in radius \citep{Gehrels:2004}.
Thus, we set the following requirements for the optical design of KOOLS--IFU:
(1) The total FoV is wider than \timeform{20''} to catch the target.
(2) The diameter ratio of a fiber core ($d_{\rm core}$) to a cladding ($d_{\rm clad}$) is equal to or larger than 0.8.
The fiber core-to-cladding ratio gives the fiber core filling factor in the focal plane.
The filling factor is equal to 90.7\% $\times (d_{\rm core} / d_{\rm clad})^2$, where 90.7\% refers to the filling factor of close-packed circles in a plane.
(3) The spectral resolving power, $R$, is $R$ = $\lambda$ / $\Delta\lambda \gtrsim 600$.

For the input part of the fiber bundle, we chose the following parameters of the IFU optics to satisfy the conditions mentioned above.
The total FoV diameter of a hexagonally packed array is approximately 
$n_{\rm f,dia} d_{\rm clad} / (m f_{\rm tel})$ radian $= 2.1 \times 10^5 \times n_{\rm f,dia} d_{\rm clad} / (m f_{\rm tel})$ arcsec, where $n_{\rm f,dia}$ is the number of fibers on the diagonal line in the total FoV, $m$ is the magnification factor of the fore-optics of the input part, and $f_{\rm tel}$ is the telescope focal length.
We set $n_{\rm f,dia} = 13$ (127 fibers in total), $d_{\rm clad} = 125$ \micron, and $m = 1 / 3$.
These settings made the FoV equal to \timeform{29''.7}. 
To achieve the large filling factor of fiber cores in the focal plane, fibers with a 100-\micron\, diameter core were selected, and this resulted in a filling factor of 58\%.
The output part of the fiber bundle makes the 1D fiber array.
We put microlens arrays (MLAs) at the ends of the fibers to make an adequate image size of the fiber on the CCD. 
Since we use the existing collimator, camera, and grisms, the image size of the fiber cores magnified by the MLAs determines the spectral resolution.
If we use KOOLS, the spectral resolving power of the Nos. 5 and 2 grisms with \timeform{1''.8} (297 \micron) slit are 400--600 and 600--800, respectively.
The focal length and the thickness of the MLAs are 1.0 mm and 0.9 mm, respectively, and each lens of the MLAs magnifies the fiber core image $\sim$2.25 times, i.e., $\sim$225 \micron.
This value is smaller than the \timeform{1''.8} slit width, and thus the spectral resolving power of KOOLS--IFU is expected to be comparable with or better than that obtained with the \timeform{1''.8} slit if the aberration is small.

\subsection{IFU}
\label{sec:inst-details}

\subsubsection{Input Part}

\begin{figure}
\begin{center}
\includegraphics[width=160mm]{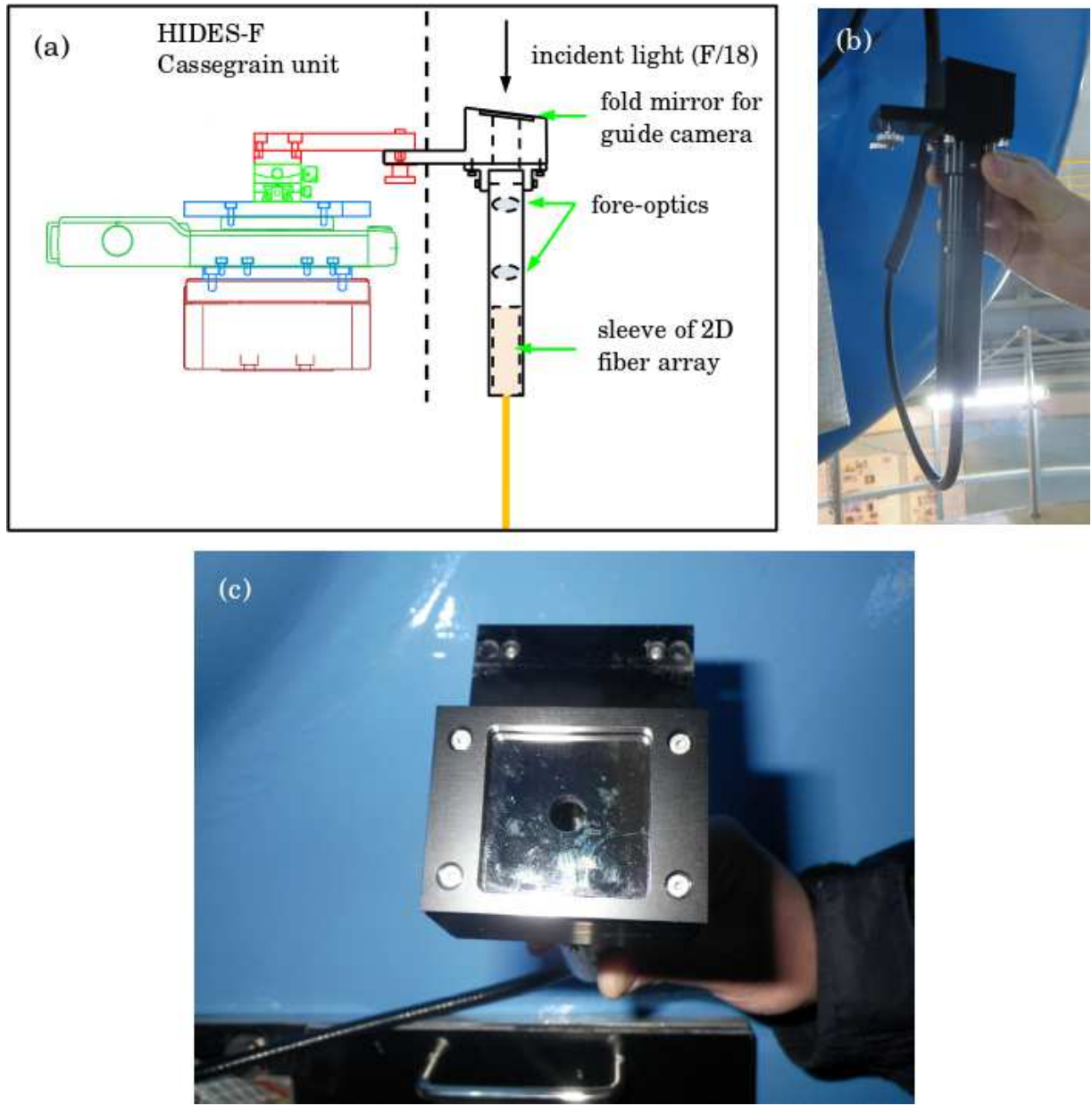}
\end{center}
\caption{(a) Conceptual layout of the input part of the IFU.
The left and right sides of this panel show the fiber mount of the HIDES-F Cassegrain unit and the IFU input part, respectively.
The HIDES-F and KOOLS--IFU input parts move perpendicular to the paper plane.
Telescope light comes from above, and its focal plane is located at the top side of the fold mirror for the guide camera.
(b) Picture of the IFU input part viewed from almost the same angle as the panel (a).
(c) Picture of the IFU input part viewed from above of the panel (a).
A mirror with a hole at its center is seen.
(Color online)
}
\label{fig:fiber-input-part}
\end{figure}

The IFU input part consists of a fold mirror, a fore-optics, and a sleeve of the 2D fiber array of the fiber bundle.
Figure \ref{fig:fiber-input-part}a shows the conceptual layout of the input part.
The fold mirror passes through the focal plane of the telescope and is tilted at \timeform{7D} to the focal plane to send light to the HIDES-F guider camera.
The diameter of the hole at the center of the mirror (seen in figure \ref{fig:fiber-input-part}c) is 8 mm, which corresponds to \timeform{49''} at the focal plane.
The size is taken to be larger than the FoV of the 2D fiber array (\timeform{30''}) so as not to block light to the fiber.
The fore-optics consists of two achromatic lenses.
Their diameters are 12.5 mm and 9 mm, and their focal lengths are 45 mm and 15 mm, respectively.
The fore-optics reduces the sky image size by 1/3 and changes the telescope F-ratio from 18 to 6.
The sleeve of the 2D fiber array is attached to the end of the cylinder of the input part.

The IFU input part is mounted on the HIDES-F Cassegrain unit.
The Cassegrain unit has a motorized XY-axis stage with three arms, and the input parts of KOOLS--IFU and HIDES-F are attached to them.
We can switch the observation instrument from HIDES-F to KOOLS--IFU in $\sim$30 seconds by moving the stage perpendicular to the paper plane of figure \ref{fig:fiber-input-part}a.
The position angle of the KOOLS--IFU FoV is constant because the HIDES-F Cassegrain unit is mounted on the equatorial telescope at a fixed position angle.

The HIDES-F calibration unit is used for the wavelength calibration of KOOLS--IFU.
Th-Ar light from the calibration unit is led to the Cassegrain focus by a retractable mirror (see figure 3 of \cite{Kambe:2013}).
Because the spot size of the calibration source (\timeform{2''.4}) is much smaller than the total FoV of KOOLS--IFU (\timeform{30''}) for obtaining calibration frames, the IFU input part is moved back and forth during the CCD exposure.
The HIDES-F guider camera is used in KOOLS--IFU observations as an offset guider camera.
The expected FoV of the guider camera is $\sim$\timeform{3'} although the FoV of the center (\timeform{49''} diameter) is not available because there is a hole at the center of the fold mirror of the IFU input part.
We confirmed that the difference between the focus positions of the HIDES-F guider camera and the KOOLS--IFU at the fiber ends of the 2D array was smaller than 0.05 mm away from the secondary mirror position of the telescope.
Hence, focusing at the IFU input part can be performed by using the guider images.

\subsubsection{Fiber Bundle}
\label{sec:inst-fiberbundle}

\begin{figure}
\begin{center}
\includegraphics[width=160mm]{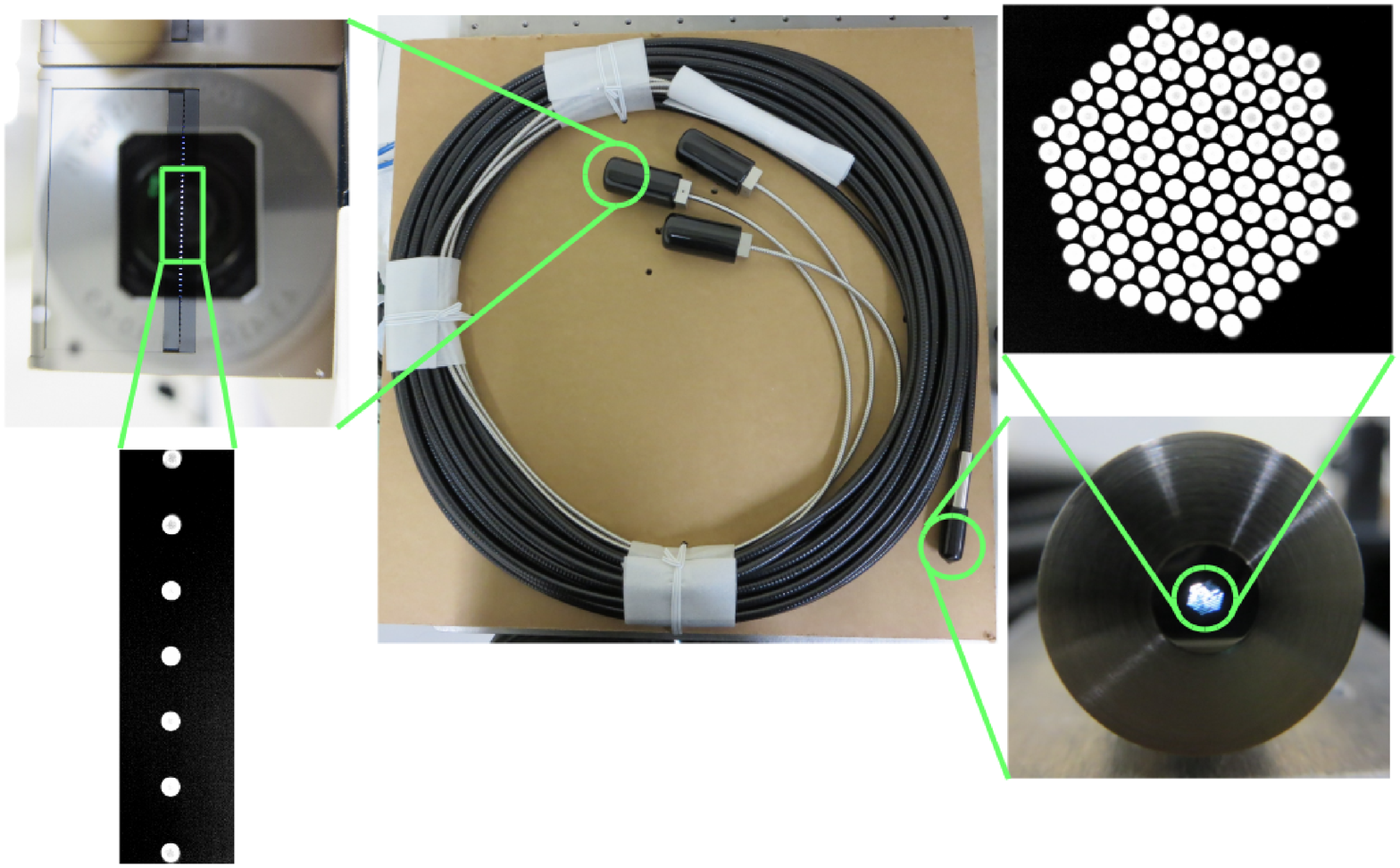}
\end{center}
\caption{Pictures of the KOOLS--IFU fiber bundle.
The center panel shows the overall view of the fiber bundle. 
The fibers are bundled in one branch and covered by a stainless tube with black polyvinyl chloride, while the last 1 m of the fibers on the 1D array side are branched into three bundles and covered by stainless tubes.
In the center panel, the sleeves are covered by black rubber caps.
Left and right panels show the end facets of 1D and 2D arrays, respectively.
(Color online)
}
\label{fig:fiberbundle-whole}
\end{figure}

\begin{figure}
\begin{center}
\includegraphics[width=160mm]{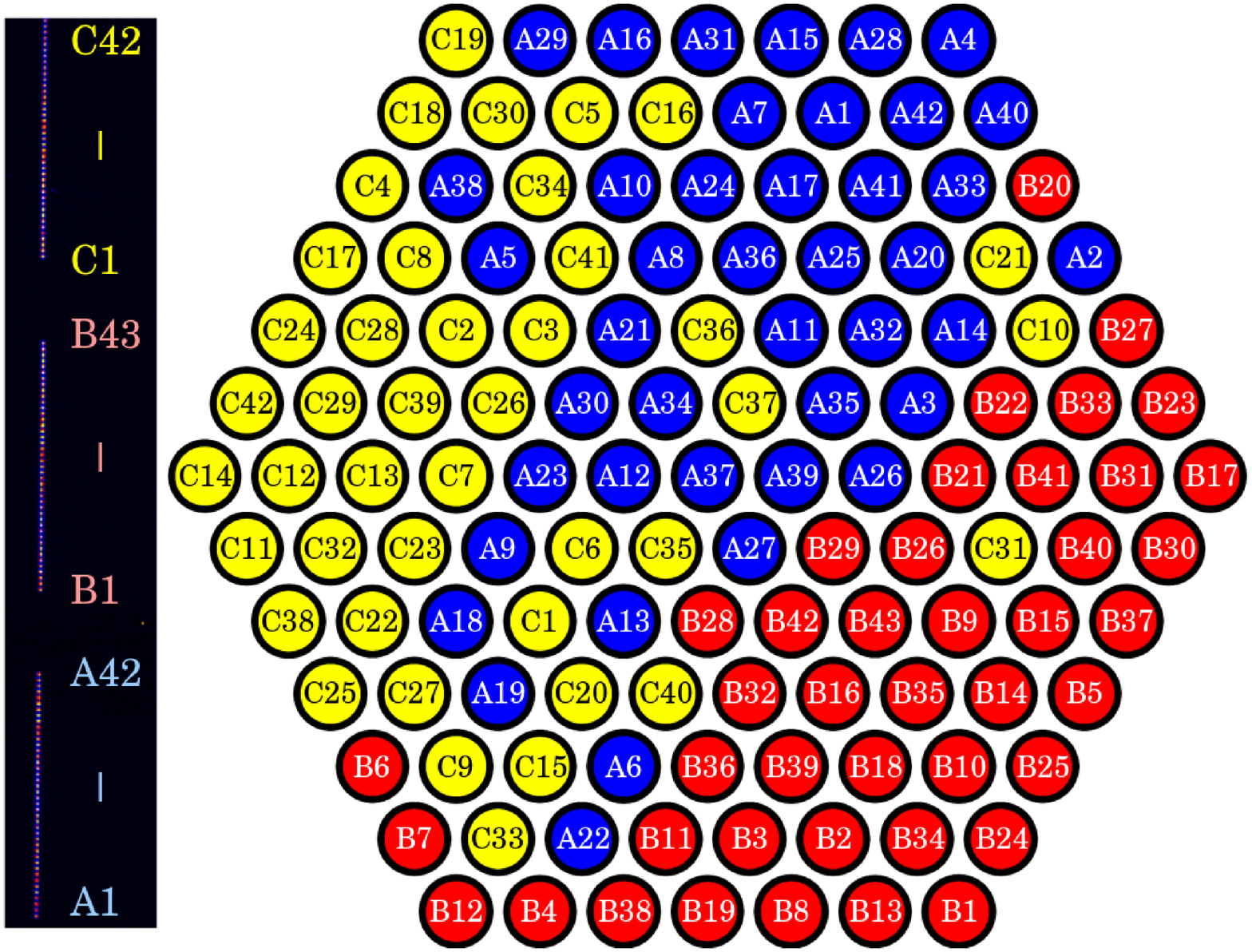}
\end{center}
\caption{Correspondence table of the fiber ends of 1D and 2D arrays.
The left panel shows the image of the fiber ends of the 1D fiber arrays on the CCD.
We numbered them in order from bottom to top as A1, A2, \ldots, A42, B1, B2, \ldots, B43, C1, C2, \ldots, C42.
The right panel shows the sketch of the fiber ends of the 2D array with the fiber IDs.
(Color online)
}
\label{fig:fiber-1d-2d-ID}
\end{figure}

Figure \ref{fig:fiberbundle-whole} shows the overall picture of the KOOLS--IFU fiber bundle, and table \ref{tb:param-fiberbundle} summarizes the parameters of the fiber bundle.
The fiber bundle was produced by Mitsubishi Cable Industries, LTD.
The fiber used in this bundle is ST100A(12); the fiber core and cladding diameters are 100 \micron\, and 125 \micron, respectively, and the fiber NA is 0.12 (F/4.2).
127 fibers are packed in this bundle.
The length of each fiber is 24 m.
The fibers are bundled in one branch, while the last 1 m of the fibers of the 1D array ends are branched into three bundles.
The fiber ends of the 2D array are packed hexagonally, and those of the 1D arrays are aligned in a line.
The correspondence table of the fiber ends of the 2D and 1D arrays is shown in figure \ref{fig:fiber-1d-2d-ID}.
The fiber bundle except for $\sim$3 m on both sides is always fixed on the telescope or other building structures.
When the input and output parts are detached, they are stored in clean boxes.
The expected FoV of a fiber is 100 $\micron$ / (1880 mm $\times$ 6) = 8.87 $\times 10^{-6}$ radian = \timeform{1''.83} and the expected total FoV of the fiber bundle is about 13 $\times$ 125 \micron\, / (1880 mm $\times$ 6) = \timeform{29''.7}.
The fiber pitch of 1D arrays is 353.55 \micron\, (= 250 \micron\, $\times \sqrt{2}$).

\begin{table}
\tbl{Basic parameters of the fiber bundle and MLAs}{
\begin{tabular}{cc}
\hline
Fiber core/cladding diameter & 100/125 \micron \\
Fiber NA & 0.12 \\
Length of each fiber & 24 m \\
Number of fibers & 127 \\
Fiber end arrangement of 2D array & hexagonal close-packed \\
Fiber pitch of 1D array & 353.55 \micron \\
MLA material & fused silica \\
Radius of curvature of MLA lenses & 0.47 mm \\
MLA lens pitch & 250 \micron \\
\hline
\end{tabular}}
\label{tb:param-fiberbundle}
\begin{tabnote}
\end{tabnote}
\end{table}

\subsubsection{Output Part}
\label{sec:inst-output}

\begin{figure}
\begin{center}
\includegraphics[width=160mm]{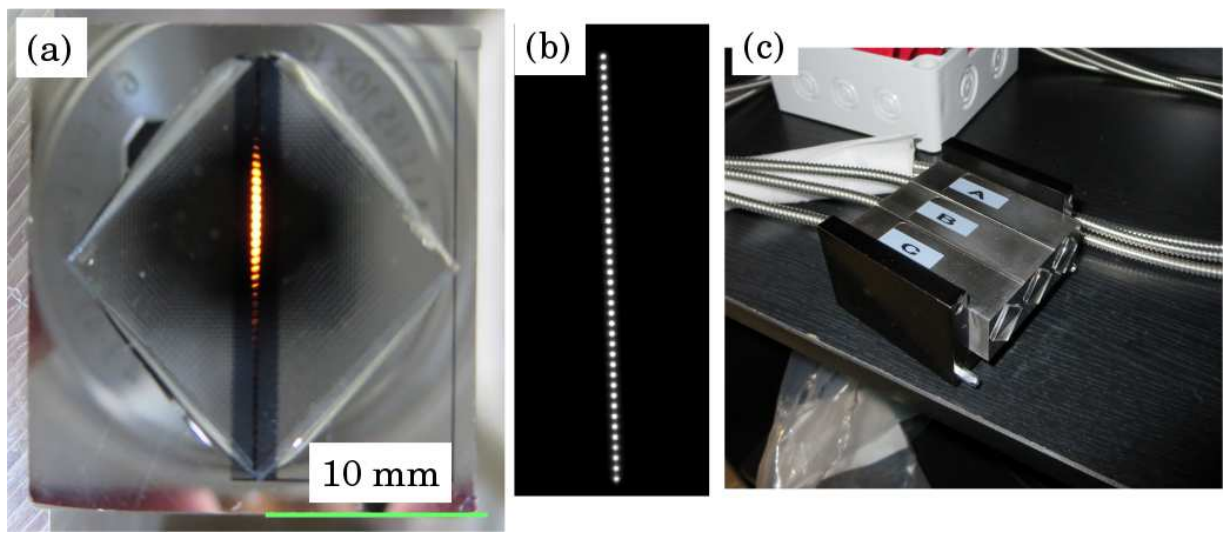}
\end{center}
\caption{(a) MLA-pasted fiber end facet of a 1D array of the fiber bundle with lightening the 2D fiber end.
The fiber ends in this sleeve are aligned in a line vertically in this panel.
Lenses on one diagonal line in a quadratic grid are matched with the fiber ends.
Light from only upper-middle fibers is visible because MLA lenses collimate light from the fibers (F $\sim$ 9) and the camera taking this picture captured light from these fibers.
(b) Same as the panel (a) but taken in a dark room.
MLA lenses magnify the fiber end images (see the lower-left panel of figure \ref{fig:fiberbundle-whole}).
(c) Output part of the IFU.
The sleeves of the 1D fiber arrays are fixed to a U-shaped block.
It is mounted inside KOOLS as a unit.
(Color online)
}
\label{fig:fiberbundle-1d-mla}
\end{figure}

We pasted the MLAs on the sleeve faces of the fiber ends of the 1D arrays with ultraviolet curable resin to change the F-ratio of the fiber output beams (figure \ref{fig:fiberbundle-1d-mla}).
The MLAs are made of fused silica, their dimensions are 12 mm $\times$ 12 mm $\times$ 0.9 mm, the radius of curvature of a lens is 0.47 mm, and a lens aperture is a 225 \micron-diameter circle.
Although the lens packing of the MLA is a quadratic grid, the lenses are matched with the fiber ends in only a diagonal line.
An MLA magnifies the fiber end images by $\sim$2.25 times, and thus the diameter of the fiber-end image is $\sim$225 \micron.
If the F-ratio of the light from the fiber is 4.2, then the F-ratio after the MLA is calculated to be $\sim$9.
The expected diameters of the fiber-end images and the F-ratio after the MLAs fulfill the specification mentioned in section \ref{sec:optical-design}.

All of the sleeves of the 1D fiber arrays are fixed to a U-shaped block (figure \ref{fig:fiberbundle-1d-mla}c) and mounted inside KOOLS as the IFU output part (figure \ref{fig:kools-layout-fiber}).
KOOLS is placed on the floor of the telescope dome during KOOLS--IFU observations.
The fiber core images enlarged by MLAs are focused at the CCD by the camera lenses.
The reduced sizes of the fiber core images (60.8 \micron\, = 225 \micron\, $\times$ 0.27) and fiber pitches (95.5 \micron) correspond to 4.1 and 6.4 pixels without binning, respectively.
We usually read the CCD using 2 $\times$ 2 on-chip binning mode to reduce the effect of readout noise.

\subsection{Expected Throughput}
\label{sec:inst-throughput}

Table \ref{tb:expected-throughput-IFU} shows the expected maximum throughput of IFU components. 
If all the flux comes through a fiber core, the expected IFU throughput is calculated to be 79\%.
Next, including the spectrograph of KOOLS, we estimate the KOOLS--IFU throughput.
The peak throughput of KOOLS is about 5\%, 10\%, 10\%, and 15\% with the No. 5, No. 2, VPH 495, and VPH 683 grisms, respectively\footnote{$<$http://www.oao.nao.ac.jp/cgi-bin/kools\_etc1.cgi$>$}.
However, the aluminum fold mirror at the center of figure \ref{fig:kools-layout-fiber} is not used with KOOLS--IFU.
By considering this effect, the expected peak throughput of KOOLS--IFU is 4.3\%, 8.7\%, 8.7\%, and 13\% with the No. 5, No. 2, VPH 495, and VPH 683 grisms, respectively, though some fibers suffer from a small amount of vignetting (corresponding to the throughput of $\sim0.98$) due to the KOOLS optics.
It should be noted that the effective throughput is worse than the above values due to the filling factor of fiber cores at the 2D array.  
The effective throughput depends on a seeing size, and we come back to this point below.

\begin{table}
\tbl{Expected maximum throughput of IFU}{
\begin{tabular}{ccc}
\hline
Optical component & Number of surface & Throughput per surface \\
\hline
Cover glass at HIDES-F Cassegrain unit & 2 & 0.99 \\
F-converting lenses & 4 & 0.985 \\
Input fiber surface & 1 & 0.96 \\
Fiber transmission$^a$ & 1 & 0.95 \\
Output MLA aperture size & 1 & 0.95 \\
Output MLA lens surface & 1 & 0.99 \\
\hline
Total & & 0.79 \\
\hline
\end{tabular}}
\label{tb:expected-throughput-IFU}
\begin{tabnote}
a: The fiber length is 24 m, and the wavelength of light is 6500 \AA.
\end{tabnote}
\end{table}

\section{Test Observations and Data Reduction}
\label{sec:test-obs}

To measure the actual parameters of KOOLS--IFU, such as the fiber image sizes on the CCD, the total FoV, and sensitivity, we carried out daytime tests and test observations mounted on the OAO 188-cm telescope.
Almost all frames of astronomical objects used in this paper were obtained on 2014 October 10 or 14, or 2014 December 26 or 27.
The weather was fine on October 14 and December 27, partly cloudy on December 26, and almost cloudy on October 10.
Seeing was \timeform{1''.5}--\timeform{2''.5}, \timeform{1''.5}--\timeform{2''.0}, and \timeform{1''.5}--\timeform{3''.5} on October 14, December 26, and December 27, respectively.

General reduction processes for fiber spectroscopic data frames were performed: bias subtraction, spectrum extraction, flat fielding, wavelength calibration, sky subtraction, and flux calibration.
We used IRAF and the original software for KOOLS--IFU data for data reduction.
The IRAF imred.hydra package \citep{Barden:1994, Barden:1995} was used for spectrum extraction, flat fielding, and wavelength calibration.
For wavelength calibration, a Th-Ar lamp for the HIDES-F and sky emission lines in object frames were used.
The sky background spectrum was made by combining spectra of the sky region surrounding the target object where the object flux was not detected.
The flux loss at the 2D fiber array due to the filling factor of fiber cores depends on seeing and the object position on the 2D array, and differs from frame to frame.
This is because KOOLS--IFU is spatially under-sampling.
The fiber pitch at the 2D fiber array and typical seeing at OAO are \timeform{2''.3} and \timeform{1''}--\timeform{2''} respectively.
We estimated the flux loss for a point source from flux ratios of the seven fibers of the brightest fiber and around it, comparing them with those calculated from models of a point source with a Gaussian profile at various positions and FWHMs.

\section{Performance of KOOLS--IFU}
\label{sec:result}

\subsection{Image Quality}
\label{sec:image-quality}

\begin{figure}
\begin{center}
\includegraphics[width=160mm]{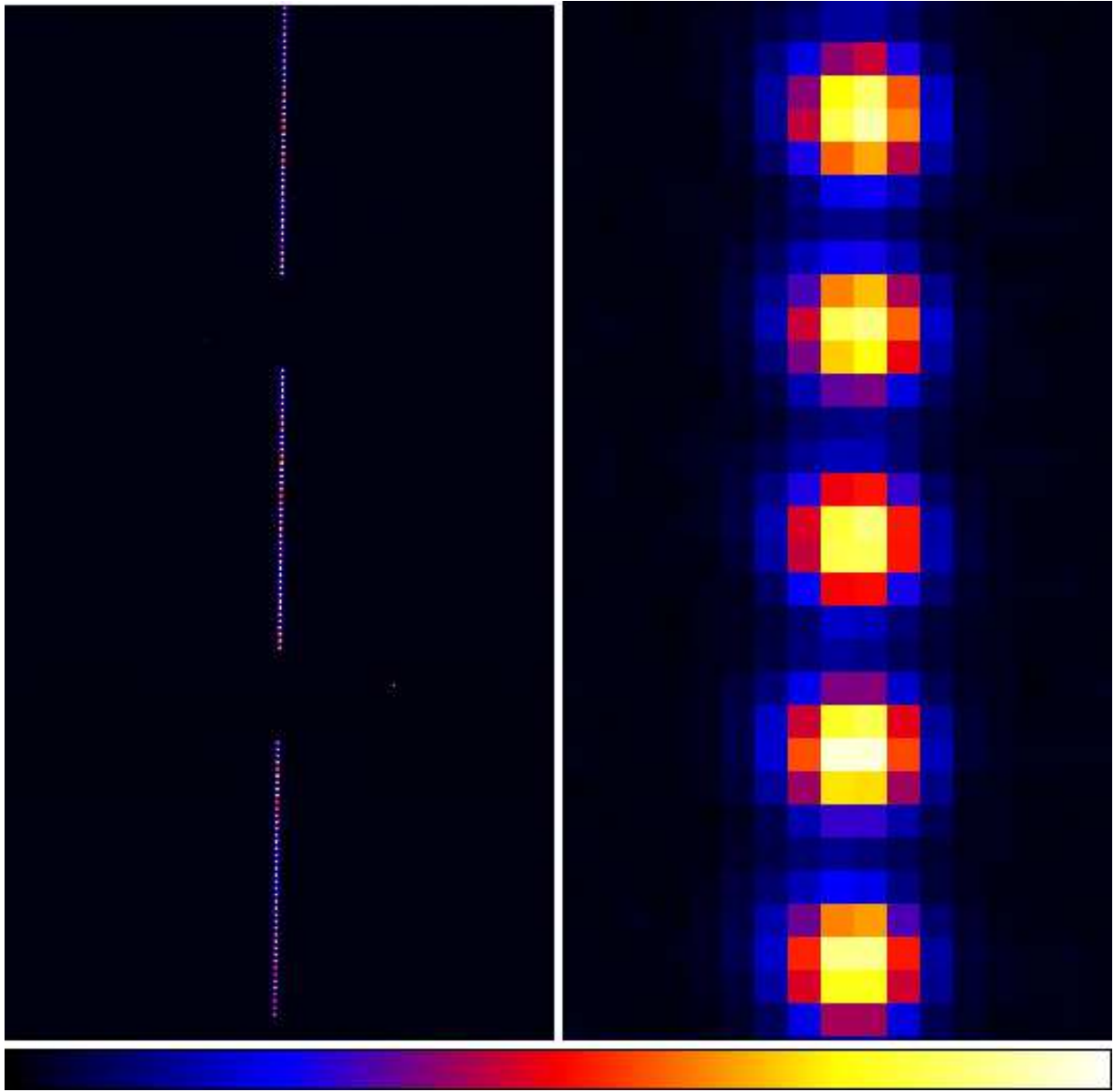}
\end{center}
\caption{(left) Whole and (right) enlarged images of the fiber ends of the 1D arrays of the fiber bundle taken by the CCD.
This image was obtained in 1 $\times$ 1 binning mode.
(Color online)
}
\label{fig:fiber-1d-whole-zoom}
\end{figure}

\begin{figure}
\begin{center}
\includegraphics[width=160mm]{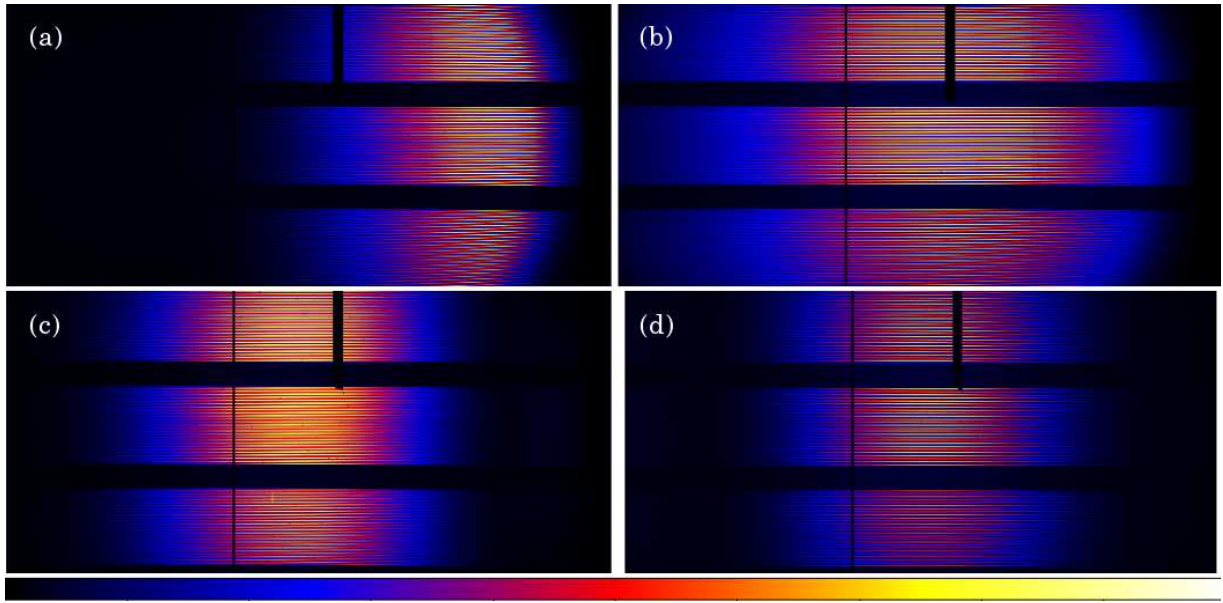}
\end{center}
\caption{Flat frames with (a) No. 5, (b) No. 2, (c) VPH 495, and (d) VPH 683 grisms.
The order sorting filter Y49 was used for the No. 2 and VPH 683 frames.
Red light goes to the right side of the frames.
(Color online)
}
\label{fig:flat-all-grism}
\end{figure}

Figure \ref{fig:fiber-1d-whole-zoom} shows the fiber-end image of the 1D arrays obtained with the CCD in 1 $\times$ 1 binning mode during a daytime test.
The measured fiber pitch is 6.4 pixels, which agrees with the estimation in section \ref{sec:inst-output}.
The fiber image sizes are 3.1--3.4 pixels in FWHM.
The separation between neighboring fiber spectra in CCD is good enough, and the spectrum from each fiber can be easily extracted. 
The crosstalk from the neighboring fiber is $\sim$1\% at the center of the CCD and $\sim$3\% near the edges of the frames.
The positional repeatability is fairly fine as the position variations of fiber images became smaller than 0.2 pixel when we remounted the output part of the IFU on KOOLS.
Figure \ref{fig:flat-all-grism} displays the flat frames for all the grisms.
In the spectra taken with C1--C42 fibers, a wavelength range (middle of the spectra in the upper part on the CCD) is not seen due to CCD defect.
The spectra taken with VPH 495 and VPH 683 grisms shift upward on the CCD relative to the spectra taken with the Nos. 5 and 2 grisms.
The spectra from the upper four fibers (fiber IDs: C39--C42) are missed.

\subsection{Relative Throughput}
\label{sec:rel-throughput}

\begin{figure}
\begin{center}
\includegraphics[width=160mm]{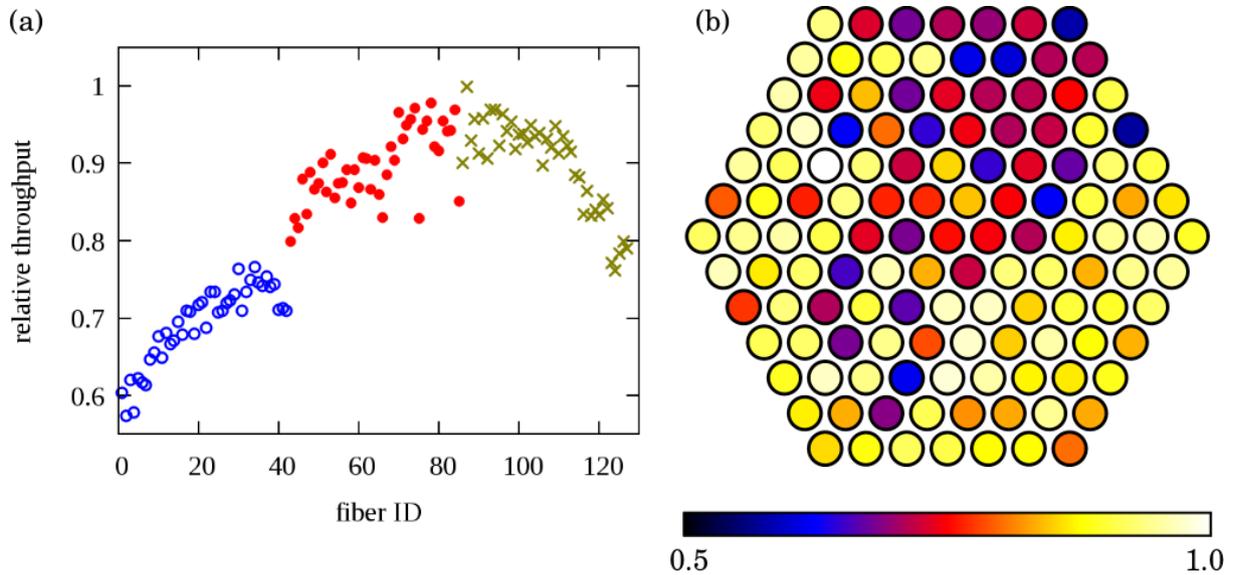}
\end{center}
\caption{Relative throughput among the fibers measured from a flat frame with No. 2 grism.
The highest value is normalized to 1.
(a) Relative throughput displayed in the format of the 1D fiber arrays.
Fiber IDs correspond to A1, A2, \ldots, A42, B1, B2, \ldots, B43, C1, C2, \ldots, C42 in order from left to right (see Fig. \ref{fig:fiber-1d-2d-ID}).
Open circles, filled circles, and crosses indicate A1--A42, B1--B43, and C1--C42 fibers, respectively.
(b) Relative throughput displayed in the format of the 2D fiber array.
The color scale is 0.5--1.0.
(Color online)
}
\label{fig:flat-no2-relative}
\end{figure}

We measured the relative throughput among the fibers using a dome flat frame with the No. 2 grism.
The flat frame was taken by pointing the telescope to a flat plate attached to the dome, which was illuminated by a halogen lamp in the back of the telescope's secondary mirror. 
Figure \ref{fig:flat-no2-relative} shows the relative throughput among the fibers.
The throughput decreases toward the edge of the IFU output part, especially through the A1--A42 fibers.
Vignetting at the filter wheels is the main cause.
Considering the relative throughput and CCD defect, we recommend using fibers at the lower-right region of the 2D array as shown in figure \ref{fig:flat-no2-relative}b to obtain spectra of point sources. 
The average throughput of the fibers relative to the highest throughput fiber was 0.835.
Almost the same result was obtained with a dome flat frame using the No. 5 grism.

\subsection{FoV and Position Angle}

We measured the FoV of KOOLS--IFU and the position angle using a frame of the binary star, $\eta$ Cas.
The observation was performed on 2014 October 10.
The expected separation and the position angle of the binary star at that time were \timeform{13''.34} and \timeform{323D.6}, respectively, i.e., the companion star was located northwest of the main star\footnote{$<$http://www.usno.navy.mil/USNO/astrometry/optical-IR-prod/wds/orb6$>$}.
The No. 2 grism was used to avoid CCD saturation, and the exposure time was 60 s.

\begin{figure}
\begin{center}
\includegraphics[width=160mm]{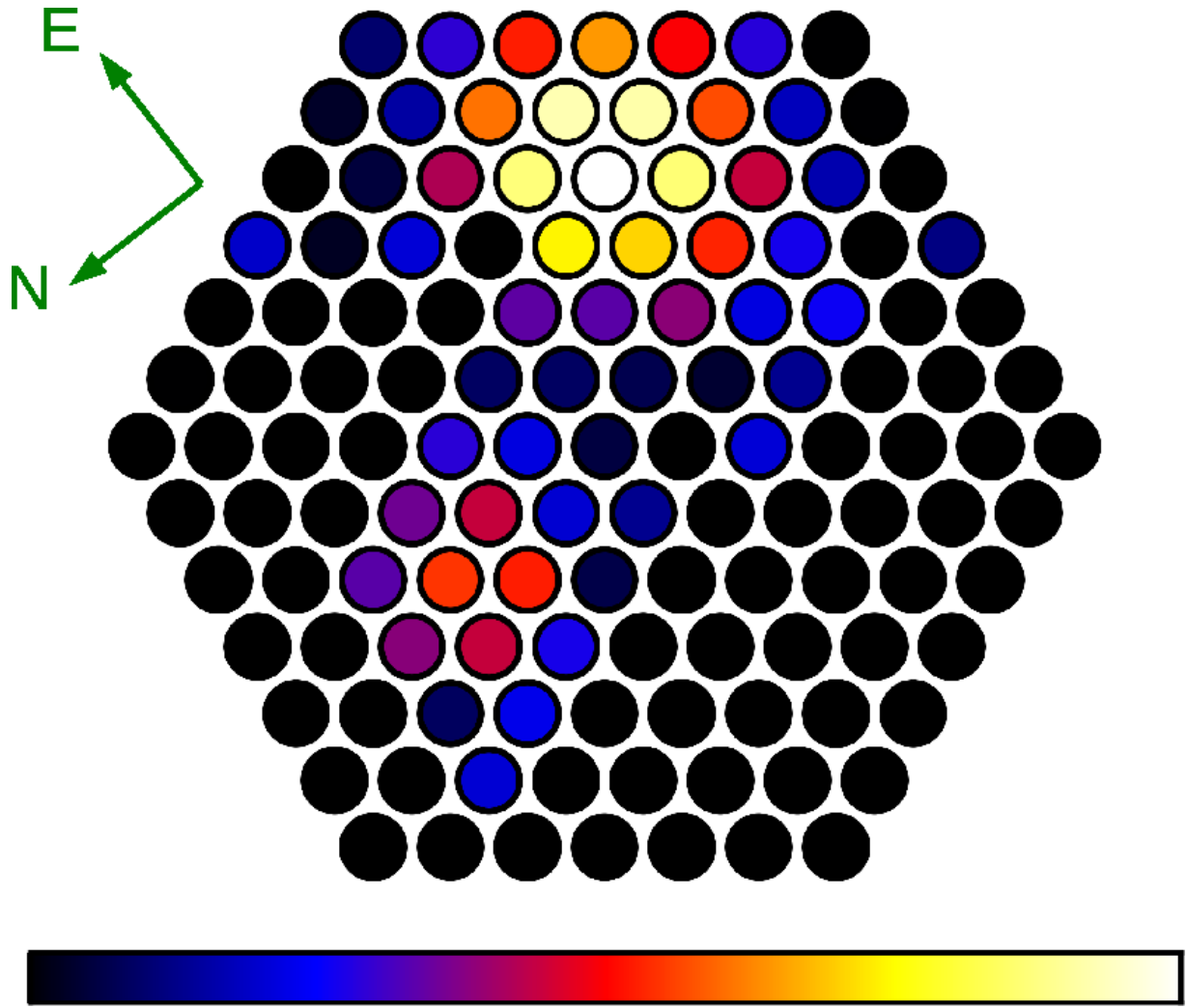}
\end{center}
\caption{Reconstructed $\eta$ Cas image obtained by KOOLS--IFU.
The intensity scale is logarithmic.
The north and east directions estimated from this frame are shown in green arrows.
(Color online)
}
\label{fig:image-eta-cas}
\end{figure}

Figure \ref{fig:image-eta-cas} shows a reconstructed $\eta$ Cas image using a logarithmic intensity scale.
We measured the precise positions of both stars in this image from the flux ratios of the local-brightest fiber and six fibers around it, which is the same method as estimating the flux loss at the 2D fiber array.
The centers of the stars are separated by 5.69 $\pm$ 0.14 fibers, and thus, one fiber separation corresponds to \timeform{2''.34} $\pm$ \timeform{0''.05}.
From the ratio of $d_{\rm core}$ to $d_{\rm clad}$ of 0.8, the FoV of a fiber is estimated to be \timeform{1''.87} $\pm$ \timeform{0''.04} in diameter, and the total FoV is \timeform{30''.4} $\pm$ \timeform{0''.65}.
These values are almost consistent with those estimated in section \ref{sec:inst-fiberbundle}.
The calculated position angle of KOOLS--IFU was \timeform{127D.3} $\pm$ \timeform{1D.3}.
We roughly confirmed this by measuring the moving direction of stars on the HIDES-F guider camera when we changed the telescope pointing slightly.
 
\subsection{Observable Wavelength and Spectral Resolution}
\label{sec:spec-observable}

The observable wavelength ranges were measured from flat frames.
Those of the No. 5, No. 2, VPH 495, and VPH 683 grisms were 4030--7310 \AA, 5020--8830 \AA, 4160--6000 \AA, and 6150--7930 \AA, respectively.
These ranges included the unobservable wavelength regions due to the defects on the CCD; 4044--4088 \AA\, and 6890--7024 \AA\, (No. 5), 6296--6344 \AA\, and 8568--8600 \AA\, (No. 2), 4834--4858 \AA\, (VPH 495), and 6594--6625 \AA\, (VPH 683) for all the fibers; and 4940--5028 \AA\, (No. 5), 6972--7050 \AA\, (No. 2), 5160--5194 \AA\, (VPH 495), and 7017--7062 \AA\, (VPH 683) for the C1--C42 fibers.

Tables \ref{tb:spec-resolution-no52} and \ref{tb:spec-resolution-vph} show the spectral resolution of all the grisms measured from the Th-Ar lines in the comparison frames and from the night sky lines in the object frames.
The spectral resolving power $\lambda / \Delta\lambda$ of the No. 5, No. 2, VPH 495, and VPH 683 grisms was about 400--600, 600--900, 1000--1200, and 1800--2400, respectively.
The measured values of the Nos. 5 and 2 grisms agree with those expected in section \ref{sec:optical-design}.

\begin{table}
\tbl{Spectral resolution of Nos. 5 and 2 grisms}{
\begin{tabular}{cccc}
\hline
Line & $\lambda$ (\AA) & $\Delta\lambda$ (\AA) & $\lambda / \Delta\lambda$ \\
\hline
\multicolumn{4}{c}{No. 5 grism} \\
\hline
Hg I & 4358.3 & 9.4 & 464 \\
Hg I & 5460.8 & 15.7 & 347 \\
ThAr & 6531.3 & 10.9 & 601 \\
ThAr & 6965.4 & 11.0 & 633 \\
ThAr & 7067.2 & 10.8 & 656 \\
\hline
\multicolumn{4}{c}{No. 2 grism} \\
\hline
Hg I & 5460.8 & 9.7 & 563 \\
ThAr & 6965.4 & 8.4 & 829 \\
ThAr & 7067.2 & 8.4 & 846 \\
ThAr & 7384.0 & 9.2 & 805 \\
ThAr & 7635.1 & 9.2 & 827 \\
ThAr & 7948.2 & 8.9 & 890 \\
ThAr & 8264.5 & 9.0 & 914 \\
ThAr & 8521.4 & 9.0 & 943 \\
\hline
\end{tabular}}
\label{tb:spec-resolution-no52}
\begin{tabnote}
\end{tabnote}
\end{table}

\begin{table}
\tbl{Spectral resolution of VPH 495 and VPH 683 grisms}{
\begin{tabular}{cccc}
\hline
Line & $\lambda$ (\AA) & $\Delta\lambda$ (\AA) & $\lambda / \Delta\lambda$ \\
\hline
\multicolumn{4}{c}{VPH 495 grism} \\
\hline
ThAr & 4277.5 & 4.5 & 961 \\
ThAr & 4348.1 & 3.6 & 1215 \\
ThAr & 4493.3 & 3.1 & 1452 \\
ThAr & 4704.0 & 4.1 & 1160 \\
ThAr & 4879.9 & 4.5 & 1089 \\
ThAr & 5231.2 & 4.8 & 1096 \\
\hline
\multicolumn{4}{c}{VPH 683 grism} \\
\hline
ThAr & 6457.2 & 3.6 & 1814 \\
ThAr & 6965.4 & 3.4 & 2078 \\
ThAr & 7067.2 & 3.3 & 2135 \\
ThAr & 7272.9 & 3.2 & 2264 \\
ThAr & 7384.0 & 3.1 & 2384 \\
ThAr & 7635.1 & 3.1 & 2460 \\
ThAr & 7724.2 & 3.2 & 2421 \\
\hline
\end{tabular}}
\label{tb:spec-resolution-vph}
\begin{tabnote}
\end{tabnote}
\end{table}

\subsection{Total Throughput}
\label{sec:throughput-obs}

\begin{figure}
\begin{center}
\includegraphics[width=160mm]{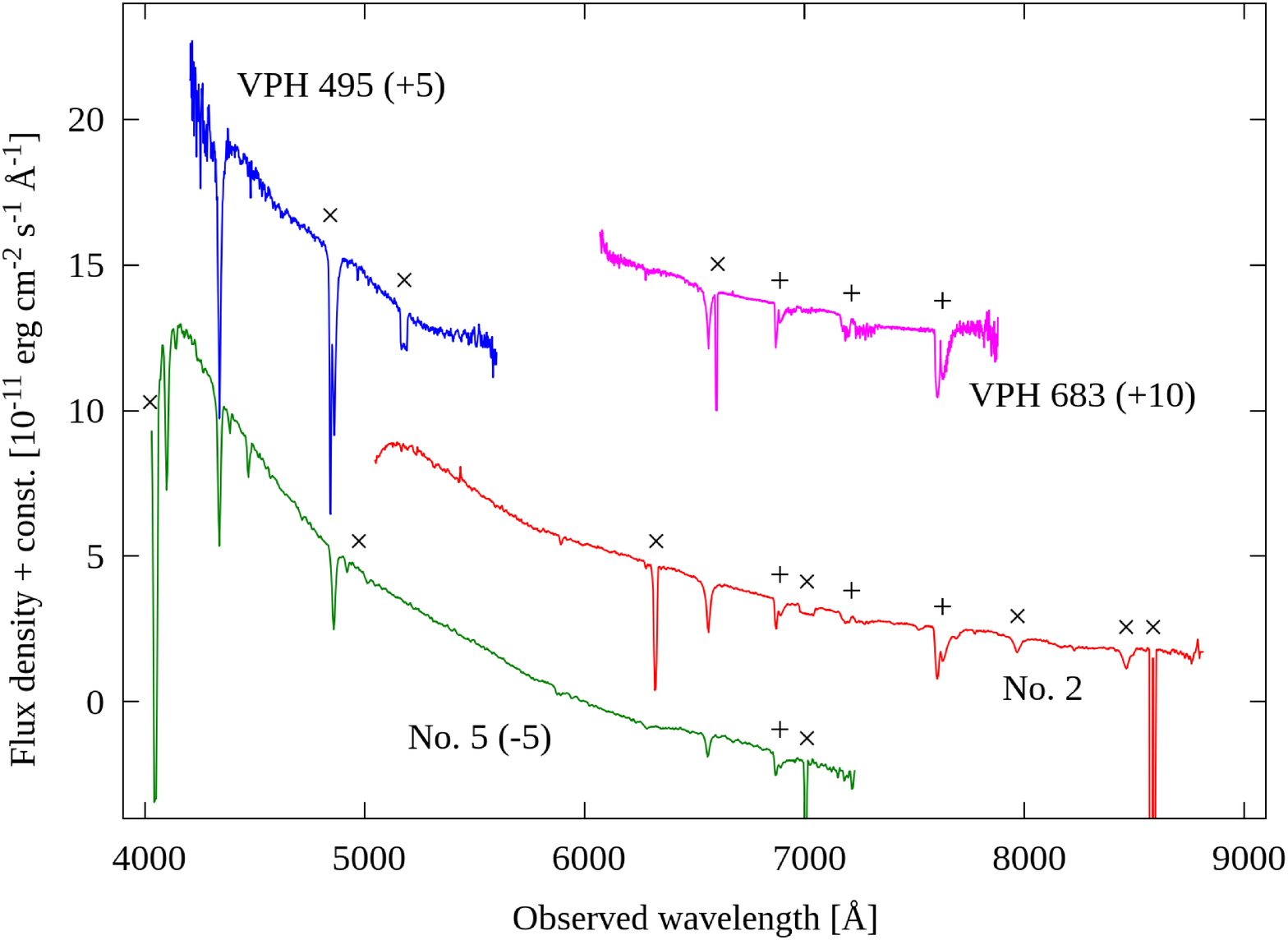}
\end{center}
\caption{Obtained spectra of spectrophotometric standard stars HD 74280 (No. 5 grism) and HD 15318 (No. 2, VPH 495, and VPH 683 grisms).
The value in the parenthesis refers to the vertical offset of each spectrum.
The features with the crosses are produced by the CCD defect mentioned in section \ref{sec:spec-observable} or second order light, while those with pluses are absorption lines by Earth's atmosphere.
(Color online)
}
\label{fig:spec-sstar}
\end{figure}

We measured the total throughput of KOOLS--IFU due to Earth's atmosphere, the telescope, IFU, KOOLS optics, and the CCD from standard star frames.
The frames of the No. 5, No. 2, and VPH 495 grisms were obtained on 2014 December 27 and that of the VPH 683 grism was obtained on 2016 February 28.
The weather was fine, and the seeing measured from the reconstructed images was \timeform{2''.1} for frames of the No. 5, No. 2, and VPH 495 grisms and \timeform{2''.6} for that of the VPH 683 grism.
Figure \ref{fig:spec-sstar} shows the obtained spectra of spectroscopic standard stars for all the grisms.
The objects were spectrophotometric standard stars HD 74280 for the No. 5 grism; and HD 15318 for the No. 2, VPH 495, and VPH 683 grisms.
The zenith distance was \timeform{31D} for the No. 5 grism, \timeform{26D} for the No. 2 and VPH 495 grisms, and \timeform{42D} for the VPH 683 grism.
While the order-sorting filter Y49 was used for the frame with the VPH 683, it was not inserted in error for the frame with the No. 2 grism.
The throughput of the No. 2 grism at wavelengths longer than $\sim$8000 \AA\, was affected by second-order light.

\begin{figure}
\begin{center}
\includegraphics[width=160mm]{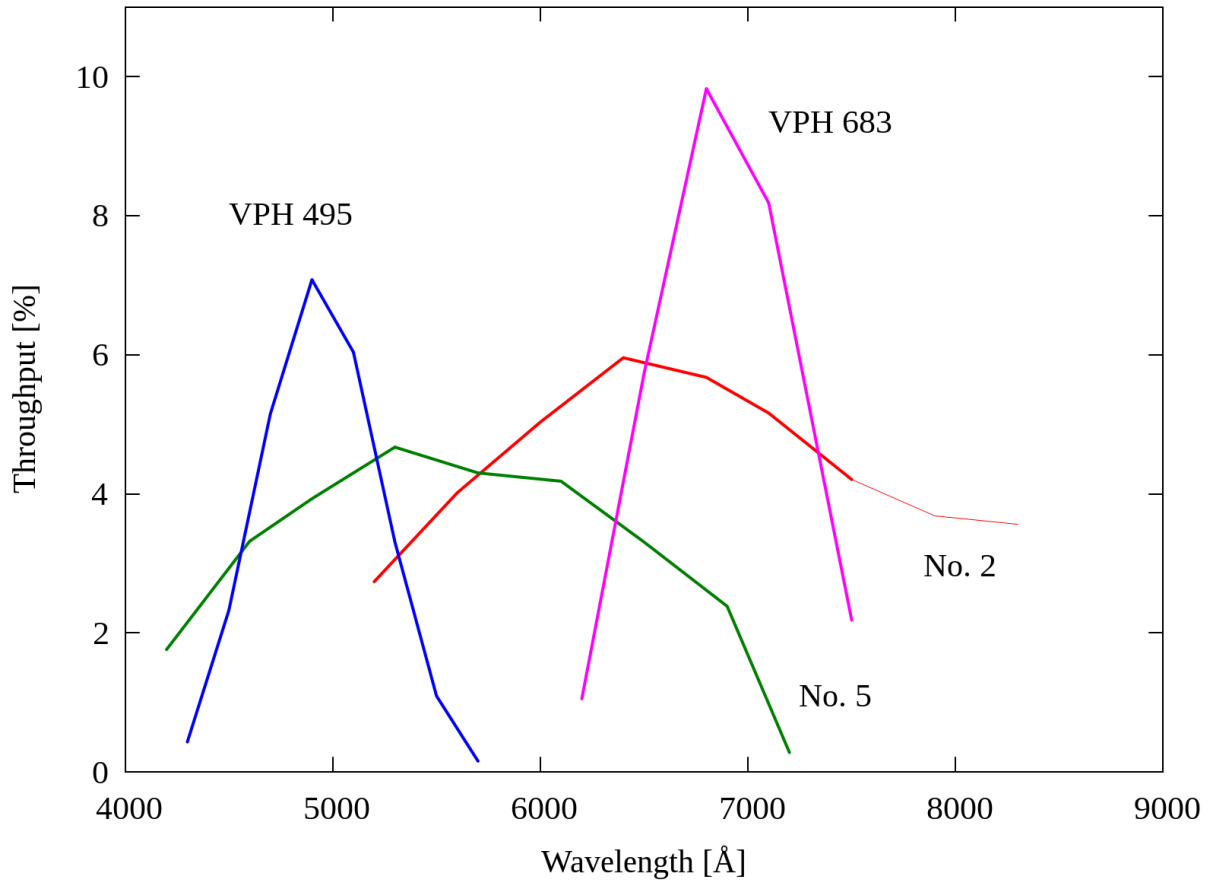}
\end{center}
\caption{KOOLS--IFU total throughput as a function of the wavelength.
Solid lines in green, red, blue, and magenta show the values of the No. 5 grism, the No. 2 grism with Y49, the VPH 495 grism, and the VPH 683 grism with Y49, respectively.
They display the throughput for the case in which all the object flux comes through the highest throughput fiber.
The throughput of the No. 2 grism at wavelengths longer than $\sim$8000 \AA\, is affected by second order light, and thus, the values are upper limit.
(Color online)
}
\label{fig:throughput-obs}
\end{figure}

If all the object flux comes through the highest throughput fiber, the peak throughput for the No. 5, No. 2, VPH 495, and VPH 683 grisms was 4.7\%, 7.1\%, 7.1\%, 3.5\%, respectively.
The throughput of the No. 5 grism was about the same as the value expected in section \ref{sec:rel-throughput}, and those of the No. 2 and VPH 495 grisms were slightly smaller.
The throughput of VPH 683 grism was about 25\% of the expected value.
After the observation, we found that the order-sorting filter Y49 was dirty, and its throughput was $\sim$30\%.
We cleaned it and its throughput recovered to 84\%.
If Y49 had been clean at the time of the observation, then we calculated that the throughput of the No. 2 and VPH 683 grisms with Y49 would have been 6.0\% and 9.8\%, respectively.
The measured total throughput is smaller by $\sim$10--30\% than those estimated in section \ref{sec:inst-throughput}.
Vignetting at the filter wheels may be the cause.
Figure \ref{fig:throughput-obs} summarizes the best throughputs as a function of the wavelength.
%It can be seen that all the object flux comes through the highest throughput fiber for all the grisms with the clean Y49 filter, namely the No. 2 and VPH 683 grisms.
It can be seen that all the object flux comes through the highest throughput fiber with the clean Y49 filter.

\subsection{Limiting Magnitude}
\label{sec:lim-mag}

Using the throughput obtained in section \ref{sec:throughput-obs}, we estimated the limiting magnitudes of KOOLS--IFU with the following conditions:
(1) The sky background emission level is 1 $\times 10^{-16}$ erg cm$^{-2}$ s$^{-1}$ arcsec$^{-2}$ \AA$^{-1}$ at all wavelengths, which is typical at OAO.
(2) The CCD readout noise is 20 electron pixel$^{-1}$. 
(3) The CCD readout mode is 2 $\times$ 2 on-chip binning.
(4) 2 wavelength-pixel flux is combined.
(5) The exposure time is 1800 s. 
(6) 90\% of the object flux comes through a fiber, which means that the object position matches the center of the highest throughput fiber, and that the point spread function is the Gaussian function with a \timeform{1''.0} FWHM.
The estimated limiting magnitudes (10$\sigma$) are 18.7, 18.6, 18.3, and 18.2 in AB mag with the No. 5, No. 2 (with Y49), VPH 495, and VPH 683 (with Y49) grisms, respectively.

To examine the above estimation, we observed the star SDSS J034959.88+362047.5 ($r^{\prime}$ = 18.5 AB mag; \cite{Alam:2015}) with the No. 2 grism but without Y49 on 2014 December 27.
The weather was fine, and the zenith angle was \timeform{2D.2}--\timeform{5D.3}.
The CCD readout mode was 2 $\times$ 2 on-chip binning, and the exposure time was 1800 s.
The continuum flux in 8 \AA\, was detected with 3.6$\sigma$ at the wavelength range of 6400--6800 \AA.
The S/N ratio of the observed frame was smaller than expected (10.8).
However, the sky condition at the time of the observation was not as good as assumed in the above sensitivity estimation, i.e., seeing was $\sim$\timeform{2''}, and the fiber relative throughput at the object (fiber ID: A38) was 0.74.
If we consider these conditions, the expected S/N ratio of an 18.5 AB mag star is 4.2, which is consistent with the observed S/N ratio.

\subsection{Guider Camera}

\begin{figure}
\begin{center}
\includegraphics[width=160mm]{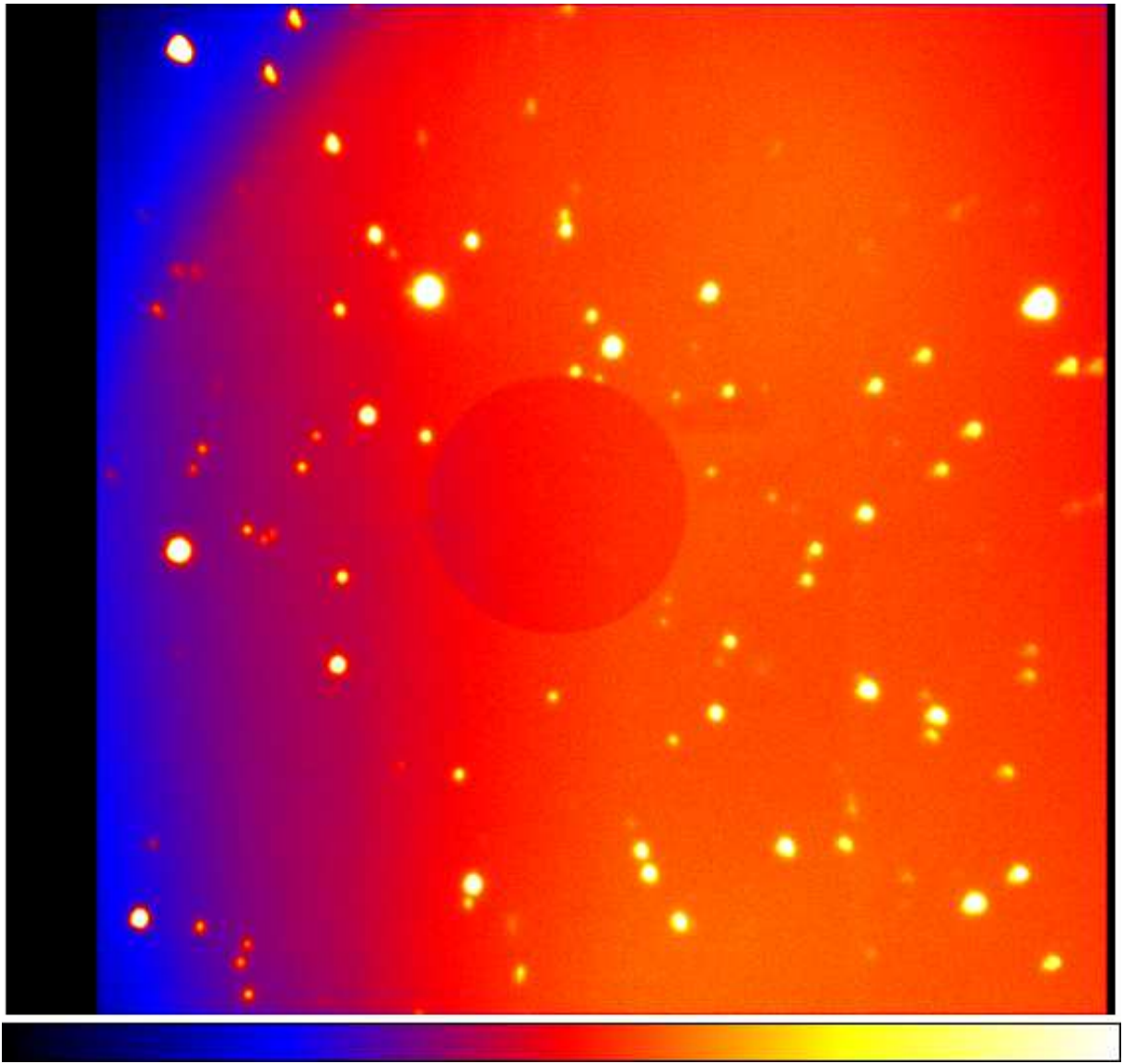}
\end{center}
\caption{Image of the open cluster NGC 1245 obtained by the HIDES-F guider camera with the wide FoV mode.
Right ascension and declination of the center of this FoV were \timeform{03h14m43s.2} and \timeform{47D15'03''}, respectively.
The FoV is \timeform{3'.24}.
Approximately north is up and west is left.
The exposure time was 15 s and no filter was used.
The background level was high due to stray light inside the guider camera system.
A faint circle around at the center is the hole at the fold mirror.
(Color online)
}
\label{fig:guider-ngc1245}
\end{figure}

The HIDES-F guider camera has both wide and narrow FoV modes \citep{Kambe:2013}.
HIDES-F uses the narrow FoV mode during exposures to check whether or not the target object comes through the HIDES-F fiber, while KOOLS--IFU uses the wide mode for the offset guide.
To obtain the parameters of the wide mode, we took an image of the open cluster NGC 1245 on 2014 December 27 (figure \ref{fig:guider-ngc1245}).
The weather was fine, and seeing was \timeform{1''.5}.
The zenith distance of the object was $\sim$\timeform{22D}.
The exposure time was 15 s, and no filter was used.
Right ascension and declination of the center of this image were \timeform{03h14m43s.2} and \timeform{47D15'03''}, respectively.
We used IRAF.daofind tasks for source detection in the guider camera image, and 114 stars were detected.
Sixty eight of detected stars were found in the USNO-B1.0 catalog\footnote{$<$http://www.nofs.navy.mil/data/fchpix/$>$} with small positional errors.
Their positions on the CCD and coordinates were fitted with linear equations.
The error was estimated from the standard error of the fitting.

The calculated pixel scale was 0.3793 $\pm$ 0.0010 (1$\sigma$) arcsec pixel$^{-1}$ in right ascension and 0.3801 $\pm$ 0.0012 arcsec pixel$^{-1}$ in declination.
The CCD had 512 $\times$ 512 pixels, and thus the total FoV of the guider camera of the wide mode was \timeform{3'.24} $\pm$ \timeform{0'.01}.
North/west was rotated by \timeform{1D.49} $\pm$ \timeform{0D.23} in a counterclockwise direction from the top/left. 
The detection limit was about 16 mag and 15 mag in Vega near and far from the FoV center, respectively, under clear skies and \timeform{1''.5} seeing.
The background level was high due to a stray light inside the guider camera system.
We covered the stray light source, and the background level lowered by a factor of $\sim$10.
The detection limit in the 2015 December run was about 17 mag in a 30-s exposure without a filter.
We estimated the expected number of stars in the FoV of the guider camera.
Even at the north Galactic pole, 0.93 star with 11--17 mag are present in the FoV \citep{Cox:2000}.

\section{Summary and Future Work}
\label{sec:summary}

To promptly obtain spectra of transients such as short GRBs and electromagnetic counterparts of GW sources, we developed an optical-fiber integral field spectrograph KOOLS--IFU and mounted it on the OAO 188-cm telescope. 
IFU consists of a fold mirror for the HIDES-F guider camera, F-converting lenses, a fiber bundle, and MLAs at the 1D fiber array.
The input part of the IFU is attached to the HIDES-F Cassegrain unit, and the output part to an existing spectrograph KOOLS.
We carried out KOOLS--IFU daytime tests and test observations and confirmed the parameters of KOOLS--IFU.
Table \ref{tb:kools-ifu-param} summarizes the KOOLS--IFU parameters.
The lag time between receiving the mock alert and starting the ToO spectroscopy was about 15 minutes but in the future this will be shortened by automating the observation processes.

\begin{table}
\tbl{KOOLS--IFU parameters}{
\begin{tabular}{ccccc}
\hline
Grism & No. 5 & No. 2$^a$ & VPH 495 & VPH 683$^a$ \\
\hline
Number of fibers & \multicolumn{4}{c}{127} \\
Spatial sampling & \multicolumn{4}{c}{\timeform{2''.34} $\pm$ \timeform{0''.05}} \\
Total FoV & \multicolumn{4}{c}{\timeform{30''.4} $\pm$ \timeform{0''.65}} \\
Wavelength coverage (\AA) & 4030--7310 & 5020--8830 & 4160--6000 & 6150--7930 \\
Spectral resolution ($\lambda/\Delta\lambda$) & 400--600 & 600--900 & 1000--1200 & 1800--2400 \\
Peak throughput (\%)$^b$ & 4.7 & 6.0 & 7.1 & 9.8 \\
Limiting magnitude (AB mag, 10$\sigma$)$^c$ & 18.7 & 18.6 & 18.3 & 18.2 \\
\hline
\end{tabular}}
\label{tb:kools-ifu-param}
\begin{tabnote}
a: With an order-sorting filter Y49\\
b: It is assumed that all the object flux comes through the highest relative throughput fiber.\\
c: The calculation conditions are written in section \ref{sec:lim-mag}.
\end{tabnote}
\end{table}

KOOLS--IFU was an open use instrument at the OAO 188-cm telescope from 2015 July to 2016 December.
We have some upgrade projects for KOOLS--IFU to improve its sensitivity.
(1) The CCD and its readout system will be replaced to increase the quantum efficiency at red wavelengths, and this in turn will reduce the readout noise level and the defect regions; and shorten the readout time.
(2) The fiber bundle with an MLA at the 2D fiber array will be improved to reduce flux loss due to the fiber filling factor.
(3) KOOLS--IFU will be mounted on the 3.8 m telescope at Okayama, Seimei telescope.
This telescope can move quickly (2--3 degrees / sec) thanks to its light-weight body \citep{Kurita:2010}, and thus is suitable to prompt spectroscopy.
These upgrades will help us obtain spectra of short GRBs, counterparts of GW sources, and spatially extended objects.

%%% Acknowledgment
\begin{ack}

We thank D. Kuroda and all the other OAO staff for the great support, especially in operating the 188 cm telescope, KOOLS, and HIDES-F.
We also thank S. Ozaki for the development of KOOLS.
This work was supported by MEXT Grant-in-Aid for Scientific Research on Innovative Areas "New Developments in Astrophysics Through Multi-Messenger Observations of Gravitational Wave Sources" (Grant Number 24103003).
This work was also supported by the Optical and Near-infrared Astronomy Inter-University Cooperation Program and Grants-in-Aid for Scientific Research (Grant Numbers 26800101, 18H05223) from the Ministry of Education, Culture, Sports, Science and Technology of Japan.

IRAF is distributed by the National Optical Astronomy Observatory, which is operated by the Association of Universities for Research in Astronomy (AURA) under a cooperative agreement with the National Science Foundation.

Funding for the Sloan Digital Sky Survey (SDSS) has been provided by the Alfred P. Sloan Foundation, the Participating Institutions, the National Aeronautics and Space Administration, the National Science Foundation, the U.S. Department of Energy, the Japanese Monbukagakusho, and the Max Planck Society. The SDSS Web site is http://www.sdss.org/.
The SDSS is managed by the Astrophysical Research Consortium (ARC) for the Participating Institutions. The Participating Institutions are The University of Chicago, Fermilab, the Institute for Advanced Study, the Japan Participation Group, The Johns Hopkins University, Los Alamos National Laboratory, the Max-Planck-Institute for Astronomy (MPIA), the Max-Planck-Institute for Astrophysics (MPA), New Mexico State University, University of Pittsburgh, Princeton University, the United States Naval Observatory, and the University of Washington.

\end{ack}

%%% References
{}

\end{document}